# The Fog of War: A Machine Learning Approach to Forecasting Weather on Mars

**Daniele Bellutta**

Undergraduate in Computer Science
The Ohio State University
bellutta.1@osu.edu

**Abstract**

For over a decade, scientists at NASA's Jet Propulsion Laboratory (JPL) have been recording measurements from the Martian surface as a part of the Mars Exploration Rovers mission. One quantity of interest has been the opacity of Mars's atmosphere for its importance in day-to-day estimations of the amount of power available to the rover from its solar arrays. This paper proposes the use of neural networks as a method for forecasting Martian atmospheric opacity that is more effective than the current empirical model. The more accurate prediction provided by these networks would allow operators at JPL to make more accurate predictions of the amount of energy available to the rover when they plan activities for coming sols.

## 1. Introduction

### 1.1. Mission Background

As a part of the ongoing Mars Exploration Rovers (MER) mission, scientists at NASA's Jet Propulsion Laboratory (JPL) must regularly monitor the Opportunity rover's health and decide what scientific activities will be performed. Much as they have done for more than a decade since Spirit and Opportunity's landing on Mars in 2004 ("Mars Exploration Rover Mission: Overview," n.d.), representatives of the MER project (both scientists and operators) periodically come together as the Science Operations Working Group (SOWG) to construct a preliminary plan of action for the rover's activities during a coming sol (Martian day) or set of sols (Mishkin and Larsen, 2006, p. 3). Though the Spirit rover's mission ended in 2011 ("Mars Exploration Rover - Spirit," n.d.), this meeting still occurs for the Opportunity rover, whose mission yet continues.

During SOWG meetings, a primary initial concern is the evaluation of the health of the rover and the restrictions placed on science activities that may be performed during a given sol. One such restriction is the amount of energy available to the rover from its solar arrays (Rayl, 2008). Along with dust covering the photovoltaic panels, the amount of sunlight at the Martian surface is an important factor in the energy constraints imposed on the rover's activities. While the scientists at JPL cannot control this aspect of the Martian environment, they must at least take it into consideration when planning activities.

To this end, operators measure the amount of sunlight reaching the rover on most sols. This quantity, which scientists at JPL denote as the Tau ($\tau$) value for a given sol, represents the atmospheric opacity at the rover's location at the time of measurement (meaning that higher values of Tau denote less sunlight reaching the surface). This is derived by performing computations on images of the Sun in the Martian sky taken using the rover's panoramic cameras (Lemmon et al., 2003, p. 3). The actual Tau value is then calculated using Beer's Law:

$$I_{observed} = I_0 e^{-\tau \eta},$$

where $I_{observed}$ is the light intensity observed from the surface of Mars, $I_0$ is the light intensity above the Martian atmosphere, and $\eta$ is the airmass (Lemmon et al., 2003, p. 3). Figure 1 demonstrates the effect of the atmospheric opacity on images taken by the rover. Once this Tau value is measured, it can be used to help estimate the amount of energy available to the rover for scientific activities.

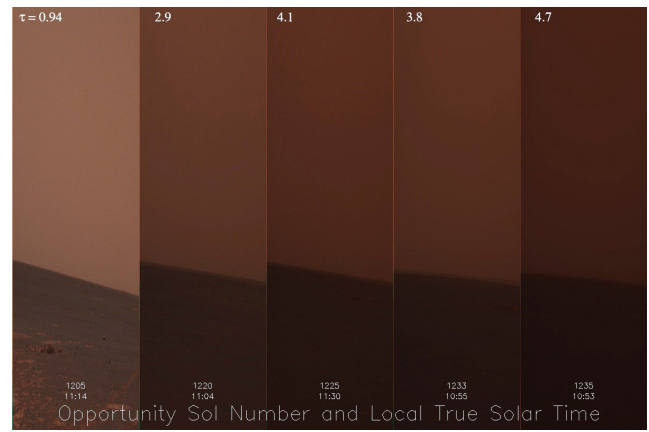

*Figure 1.* The effect of atmospheric opacity on images taken by the rovers. Image courtesy of NASA/JPL-Caltech/Cornell via "Dust Storm Time Lapse Shows Opportunity's Skies Darken," 2007.

## 1.2. The Problem

When planning the Opportunity rover's activities, one of the data products discussed during SOWG meetings is the most recently measured Tau value because of its importance for predicting how much energy will be available to the rover on subsequent sols. Currently, JPL scientists estimate future sols' Tau values manually by adding a margin to the previous day's value:

$$\tau_t = \tau_{t-1} + m,$$

where $t$ represents the sol number and $m$ is the margin (which can change from sol to sol). The constraints placed by this margin accumulate very quickly when planning operations multiple sols in advance, sharply restricting how much scientific activity can be planned.

The snowballing of the margins severely depresses the estimated amount of energy with which rover operators are allowed to work. Thus, a better method for predicting Tau values with a smaller margin of error would give more precise forecasts and therefore increase the amount of energy the rover is projected to have available. This would ultimately allow the scientists and rover drivers at JPL to plan more activities and collect more data during the remaining lifetime of the mission.

## 1.3. The (Possible) Solution

This paper outlines a new method for forecasting Tau values using neural networks. Specifically, the use of both standard and nonlinear autoregressive neural networks for this task will be discussed, and the results will be analyzed. A short but interesting experiment will follow, which involves training the networks on data from one rover (either Spirit or Opportunity) and attempting to predict the Tau values for the other rover.

## 2. Approach

### 2.1. The Data Set

Tau values measured by the Mars rovers Spirit and Opportunity are made available by the Jet Propulsion Laboratory (Lemmon, 2016). Thousands of usable data points exist from each rover, both of which have provided measurements across multiple Martian years. These values are recorded along with the sol number (the number of sols since the start of the mission) on which they were taken. Knowing when the measurement was made is paramount for being able to predict the Tau for a given sol, but having this knowledge in the form of the mission's sol number is not extremely helpful.

A suspicion that the atmospheric opacity may be tied to patterns in the Martian seasons is not entirely unreasonable. To account for this prospect, it is possible to associate the Tau measurements with the point in the Martian year at which they were taken. In order to do this, the sol numbers must be converted to the equivalent areocentric longitude ($L_s$), which represents "the longitude of the Sun as viewed from the centre of Mars" (Powell, n.d.) and consequently ties the Tau measurement to the position of Mars in its orbit around the Sun. The scientists at JPL are thoroughly acquainted with this process and therefore have resources available for this kind of preprocessing.

Once this conversion has been performed, it is possible to view the Tau data as a function of the status of the Martian year. Figure 2 shows such a visualization, and it indeed seems apparent that Tau values are at least somehow tied to the Martian seasons. Linking Tau measurements with their corresponding $L_s$ values makes them much more powerful for forecasting the atmospheric opacity in the future, since any yearly patterns can be learned and considered.

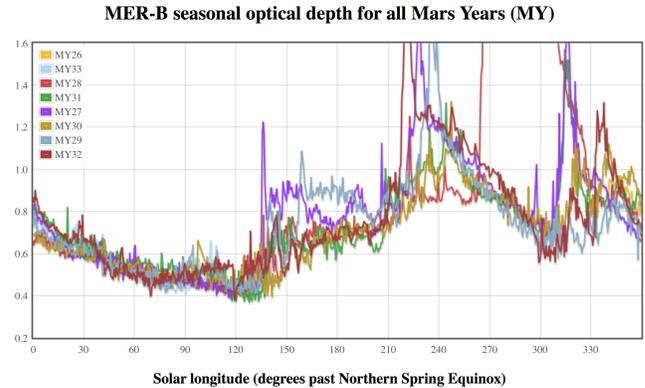

*Figure 2.* Tau measurements from Opportunity as a function of the areocentric longitudes when they were taken. Graph courtesy of Dr. Mark Lemmon, 2016.

### 2.2. The Neural Networks

Artificial neural networks have become an extremely powerful method for learning patterns between inputs and outputs (Choudhary et al., 2012, p. 3). In these networks, nodes (artificial neurons) are connected in layers with various possible arrangements. Inputs are then fed into the nodes, which use activation functions to determine their level of output to the next layer of nodes. Weights are assigned to each connection between nodes and are updated during training in order to minimize error. The output $y$ of one neuron can therefore be represented as the following:

$$y = g\left(\sum_i w_i x_i\right),$$

where $g$ is the activation function and $w_i$ is the weight for each input $x_i$.

The set of Tau data collected by the MER mission essentially represents a nonlinear time series, meaning that certain network types may be more suitable for predicting Mars's atmospheric opacity. Nonlinear autoregressive models with exogenous inputs have proven powerful for representing nonlinear time series, making them appropriate for modeling Tau data. When using such nonlinear autoregressive models, the output $y_t$ for a given point in time $t$ can be expressed as the following:

$$y_t = f\left(y_{t-1}, y_{t-2}, \ldots, y_{t-d}, x_t, x_{t-1}, \ldots, x_{t-d}\right) + e_t,$$

where $y_i$ are values in the time series being modeled for appropriate time points $i$, $x_i$ are exogenous values corresponding to the $y_i$ values from the time series, $f$ is a nonlinear function, $d$ is the number of delays to be considered by the model, and $e_t$ is the error of the model for the point in time $t$. In the specific case of modeling the atmospheric opacity on Mars, the model's outputs are Tau values, and the exogenous inputs used were the areocentric longitude $L_s$ associated with the Tau measurements (thus associating the Tau values with the Martian seasons). The nonlinear function $f$ for the models considered by this paper was a neural network.

### 2.3. Methodology

Two experiments were performed on the Tau data for this project. The first experiment involved training a nonlinear autoregressive neural network on 75% of a rover's Tau measurements (along with their corresponding $L_s$ values) using Levenberg-Marquardt backpropagation, cross-validating the network on 15% of the data, and then testing it on the rest of that rover's data. Several trials were performed for each rover: various combinations of the number of hidden nodes in the network and the number of delays considered by the model were tested. Three runs were performed for every such combination to account for variations in training, and the results outlined in this paper show the average values for the correlation coefficient and mean squared error (MSE) across the three runs.

Each combination of different numbers of delays and hidden nodes was also tested with duplicate delays added manually by including the Tau and $L_s$ values for previous sols. Though this is an included feature of autoregressive networks, the duplicate delays were added in an attempt to give more weight to more recent Tau values and their corresponding $L_s$ in order to discern if this would improve the network's predictions. The first experiment was also repeated using a standard neural network instead of an autoregressive network in order to validate its results.

For the second experiment, the first experiment was repeated at a much smaller scale and with one significant difference: though the autoregressive network was trained on the data from one rover, it was tested on the data from the other rover. This experiment was only performed using twelve hidden nodes and two delays, since the results from the first experiment (discussed later) overwhelmingly show that the number of delays and number of hidden nodes do not have much effect on the performance of the network. Three runs were still performed for each kind of input and only the means of the results across those three runs are discussed in this paper.

## 3. Results

### 3.1. First Experiment

Appendices A and B show the MSE values and correlation coefficients for training and testing a nonlinear autoregressive network on the data from each rover. When examining the full results in the appendices, it seems clear that the number of nodes and number of delays has very little, if any, deterministic effect on the performance of the network over the range of values tested for this project. Therefore, Tables 1 and 2 have been made to show only the averaged results from across the combinations of numbers of nodes and delays.

| | Averaged Nonlinear Autoregressive Network Results for Opportunity | | |
|---|---|---|---|
| Inputs | $L_{s,t}$ | $L_{s,t}, \tau_{t-1}, L_{s,t-1}$ | $L_{s,t}, \tau_{t-1}, L_{s,t-1}, \tau_{t-2}, L_{s,t-2}$ |
| MSE | 0.0034 | 0.0036 | 0.0035 |
| Correlation | 0.9863 | 0.9859 | 0.9861 |

*Table 1.* Summarized results from training and testing a nonlinear autoregressive neural network on data from Opportunity.

| | Averaged Nonlinear Autoregressive Network Results for Spirit | | |
|---|---|---|---|
| Inputs | $L_{s,t}$ | $L_{s,t}, \tau_{t-1}, L_{s,t-1}$ | $L_{s,t}, \tau_{t-1}, L_{s,t-1}, \tau_{t-2}, L_{s,t-2}$ |
| MSE | 0.0051 | 0.0053 | 0.0049 |
| Correlation | 0.9890 | 0.9889 | 0.9895 |

*Table 2.* Summarized results from training and testing a nonlinear autoregressive neural network on data from Spirit.

As can be seen from the results, attempting to give more weight to later measurements by manually adding in duplicate delays does not have a significant effect on the network's performance. It should also be noted that training and testing on the data from Spirit yields a slightly higher error rate. This is most likely due to the fact that there is less data available for Spirit (from 2004 to 2010) than for Opportunity (from 2004 to 2016).

The results from repeating the first experiment using a standard neural network (without the autoregression) are listed in full in Appendices C and D. Tables 3 and 4 summarize these results in a similar way as Tables 1 and 2. One should bear in mind that the standard neural network had no parameter for the number of delays to consider. The results for these trials do confirm the findings of the first set that used a nonlinear autoregressive network. The standard neural network initially performs poorly when only the $L_s$ for the desired sol is used as input, but once previous Tau measurements and their correspond $L_s$ values are added as inputs (which is what the autoregressive network does automatically), the results immediately improve to the same levels as when using the autoregressive network.

| Averaged Standard Network Results for Opportunity | | | |
|---|---|---|---|
| Inputs | $L_{s,t}$ | $L_{s,t}, \tau_{t-1}, L_{s,t-1}$ | $L_{s,t}, \tau_{t-1}, L_{s,t-1}, \tau_{t-2}, L_{s,t-2}$ |
| MSE | 0.0761 | 0.0035 | 0.0034 |
| Correlation | 0.6255 | 0.9859 | 0.9864 |

*Table 3.* Summarized results from training and testing a standard neural network on data from Opportunity.

| Averaged Standard Network Results for Spirit | | | |
|---|---|---|---|
| Inputs | $L_{s,t}$ | $L_{s,t}, \tau_{t-1}, L_{s,t-1}$ | $L_{s,t}, \tau_{t-1}, L_{s,t-1}, \tau_{t-2}, L_{s,t-2}$ |
| MSE | 0.1266 | 0.0052 | 0.0050 |
| Correlation | 0.6711 | 0.9886 | 0.9893 |

*Table 4.* Summarized results from training and testing a standard neural network on data from Spirit.

### 3.2. Second Experiment

During the second experiment, nonlinear autoregressive networks were trained on the data from one rover but tested on the data from the other rover. These trials were only performed on networks with twelve hidden nodes and two delays, since it was discovered in the first experiment that the number of nodes and number of delays does not have a significant effect on the network's performance for forecasting Tau values.

Though it might be expected that training on one rover and testing on the other would produce unsatisfactory results, in reality this experiment yielded surprisingly good results. While the MSE values from the second experiment are at times twice the MSE values from the first experiment, they are still below 0.01. The correlation coefficients are also well above 0.98, meaning that the model still produces more than adequate results.

| Results for Training on Opportunity and Testing on Spirit | | | |
|---|---|---|---|
| Inputs | $L_{s,t}$ | $L_{s,t}, \tau_{t-1}, L_{s,t-1}$ | $L_{s,t}, \tau_{t-1}, L_{s,t-1}, \tau_{t-2}, L_{s,t-2}$ |
| MSE | 0.0074 | 0.0072 | 0.0073 |
| Correlation | 0.9853 | 0.9851 | 0.9851 |

*Table 5.* Results for training a nonlinear autoregressive neural network on data from Opportunity and testing it on data from Spirit.

| Results for Training on Spirit and Testing on Opportunity | | | |
|---|---|---|---|
| Inputs | $L_{s,t}$ | $L_{s,t}, \tau_{t-1}, L_{s,t-1}$ | $L_{s,t}, \tau_{t-1}, L_{s,t-1}, \tau_{t-2}, L_{s,t-2}$ |
| MSE | 0.0046 | 0.0059 | 0.0049 |
| Correlation | 0.9818 | 0.9779 | 0.9812 |

*Table 6.* Results from training a nonlinear autoregressive neural network on data from Spirit and testing it on data from Opportunity.

## 4. Conclusions

### 4.1. Analysis of Results

Using neural networks seems to be an extraordinarily effective method for forecasting atmospheric opacity on Mars. Both the nonlinear autoregressive networks and the standard neural networks were able to achieve remarkably low MSE values and correlation coefficients higher than 0.98 (even surpassing 0.99 in some cases). By using delays, the autoregressive networks are able to attain very low error even when only given the $L_s$ for the desired sol. The standard neural networks need at least one past Tau value (and its corresponding $L_s$) to also be provided as input in order to perform well, but they are still able to achieve similar results to those of the autoregressive networks.

Exactly why these neural networks perform so well at Tau prediction is not certain. However, a few possible explanations can be postulated for this phenomenon. Upon examining the data directly, one notices that the Tau value does not change dramatically from sol to sol. This may be a reason that these models work so well at predicting future atmospheric opacity (along with the possibility that the atmospheric opacity is heavily associated with the Martian seasons). The results seem to support this thinking, since the performance of the standard neural networks jumps considerably immediately after adding the previous sol's Tau as an input.

As for why training on one rover and testing on the other works so well, one explanation might be that both Spirit and Opportunity are not that far from the Martian equator. Spirit landed at a latitude of 14.57 degrees south, and Opportunity landed at a latitude of 1.95 degrees south (Jet

Propulsion Laboratory, n.d.). Since their landing, the rovers have not traveled far enough to significantly affect the weather they experience. In 2015, for example, Opportunity was around 42 kilometers from its landing site ("NASA's Opportunity Mars Rover Passes Marathon Distance," 2015). The fact that there is not a large difference in the longitude of the rovers' locations may be the reason that both rovers experience such similar weather.

### 4.2. Applications

Given the importance of atmospheric opacity in determining how much energy is produced by the Mars Exploration Rovers' solar arrays, better Tau predictions would enable scientists at the Jet Propulsion Laboratory to make much more precise judgements about how much energy Opportunity will have at its disposal in coming sols. This would in turn allow them to make better decisions when planning the rover's activities.

Since the amount of energy available is such a significant limitation when planning rover activities, under- or overestimating this quantity can have extremely regrettable consequences. While overestimating the amount of energy is obviously the worse of the two options, underestimating it is also undesirable because more scientific activities could possibly have been performed with the extra energy that was not considered during planning. Hence, the more precise forecasts provided by the neural networks discussed in this paper could allow JPL scientists to plan more activities in the remaining lifetime of the MER mission and make more scientific discoveries about Mars.

## 5. Acknowledgements

This research was carried out with significant support from several people at the Jet Propulsion Laboratory in Pasadena, California. Their help with acquiring and understanding the data necessary for this project was greatly appreciated.

# Appendix A: Nonlinear Autoregressive Network Results for Opportunity

| | | 1 Delay | | 2 Delays | | 3 Delays | | 4 Delays | |
|---|---|---|---|---|---|---|---|---|---|
| | Inputs | MSE | Correlation | MSE | Correlation | MSE | Correlation | MSE | Correlation |
| 6 Nodes | $L_{s,t}$ | 0.0036 | 0.9856 | 0.0033 | 0.9868 | 0.0034 | 0.9865 | 0.0035 | 0.9862 |
| | $L_{s,t}, \tau_{t-1}, L_{s,t-1}$ | 0.0034 | 0.9865 | 0.0050 | 0.9824 | 0.0032 | 0.9870 | 0.0040 | 0.9837 |
| | $L_{s,t}, \tau_{t-1}, L_{s,t-1}, \tau_{t-2}, L_{s,t-2}$ | 0.0031 | 0.9875 | 0.0033 | 0.9868 | 0.0034 | 0.9863 | 0.0033 | 0.9868 |
| 8 Nodes | $L_{s,t}$ | 0.0035 | 0.9858 | 0.0032 | 0.9872 | 0.0035 | 0.9861 | 0.0039 | 0.9853 |
| | $L_{s,t}, \tau_{t-1}, L_{s,t-1}$ | 0.0033 | 0.9867 | 0.0033 | 0.9866 | 0.0035 | 0.9862 | 0.0035 | 0.9862 |
| | $L_{s,t}, \tau_{t-1}, L_{s,t-1}, \tau_{t-2}, L_{s,t-2}$ | 0.0032 | 0.9871 | 0.0034 | 0.9864 | 0.0033 | 0.9866 | 0.0037 | 0.9853 |
| 10 Nodes | $L_{s,t}$ | 0.0035 | 0.9860 | 0.0049 | 0.9806 | 0.0031 | 0.9877 | 0.0034 | 0.9865 |
| | $L_{s,t}, \tau_{t-1}, L_{s,t-1}$ | 0.0033 | 0.9866 | 0.0036 | 0.9856 | 0.0033 | 0.9869 | 0.0035 | 0.9862 |
| | $L_{s,t}, \tau_{t-1}, L_{s,t-1}, \tau_{t-2}, L_{s,t-2}$ | 0.0035 | 0.9859 | 0.0032 | 0.9870 | 0.0037 | 0.9859 | 0.0031 | 0.9876 |
| 12 Nodes | $L_{s,t}$ | 0.0035 | 0.9861 | 0.0031 | 0.9875 | 0.0031 | 0.9874 | 0.0034 | 0.9864 |
| | $L_{s,t}, \tau_{t-1}, L_{s,t-1}$ | 0.0032 | 0.9870 | 0.0033 | 0.9868 | 0.0030 | 0.9881 | 0.0045 | 0.9825 |
| | $L_{s,t}, \tau_{t-1}, L_{s,t-1}, \tau_{t-2}, L_{s,t-2}$ | 0.0031 | 0.9875 | 0.0033 | 0.9870 | 0.0034 | 0.9865 | 0.0035 | 0.9861 |
| 14 Nodes | $L_{s,t}$ | 0.0034 | 0.9865 | 0.0032 | 0.9873 | 0.0034 | 0.9864 | 0.0037 | 0.9857 |
| | $L_{s,t}, \tau_{t-1}, L_{s,t-1}$ | 0.0032 | 0.9871 | 0.0035 | 0.9859 | 0.0033 | 0.9867 | 0.0034 | 0.9862 |
| | $L_{s,t}, \tau_{t-1}, L_{s,t-1}, \tau_{t-2}, L_{s,t-2}$ | 0.0035 | 0.9860 | 0.0030 | 0.9880 | 0.0031 | 0.9874 | 0.0034 | 0.9863 |
| 16 Nodes | $L_{s,t}$ | 0.0033 | 0.9867 | 0.0034 | 0.9869 | 0.0035 | 0.9858 | 0.0037 | 0.9854 |
| | $L_{s,t}, \tau_{t-1}, L_{s,t-1}$ | 0.0032 | 0.9871 | 0.0034 | 0.9867 | 0.0040 | 0.9849 | 0.0050 | 0.9801 |
| | $L_{s,t}, \tau_{t-1}, L_{s,t-1}, \tau_{t-2}, L_{s,t-2}$ | 0.0029 | 0.9882 | 0.0030 | 0.9878 | 0.0038 | 0.9848 | 0.0033 | 0.9868 |
| 18 Nodes | $L_{s,t}$ | 0.0033 | 0.9867 | 0.0031 | 0.9875 | 0.0033 | 0.9868 | 0.0039 | 0.9849 |
| | $L_{s,t}, \tau_{t-1}, L_{s,t-1}$ | 0.0032 | 0.9872 | 0.0031 | 0.9878 | 0.0033 | 0.9869 | 0.0034 | 0.9865 |
| | $L_{s,t}, \tau_{t-1}, L_{s,t-1}, \tau_{t-2}, L_{s,t-2}$ | 0.0032 | 0.9873 | 0.0035 | 0.9858 | 0.0034 | 0.9866 | 0.0062 | 0.9746 |
| 20 Nodes | $L_{s,t}$ | 0.0035 | 0.9860 | 0.0032 | 0.9869 | 0.0032 | 0.9872 | 0.0031 | 0.9877 |
| | $L_{s,t}, \tau_{t-1}, L_{s,t-1}$ | 0.0040 | 0.9843 | 0.0050 | 0.9808 | 0.0033 | 0.9869 | 0.0029 | 0.9883 |
| | $L_{s,t}, \tau_{t-1}, L_{s,t-1}, \tau_{t-2}, L_{s,t-2}$ | 0.0033 | 0.9868 | 0.0038 | 0.9848 | 0.0041 | 0.9841 | 0.0038 | 0.9850 |

# Appendix B: Nonlinear Autoregressive Network Results for Spirit

| | | 1 Delay | | 2 Delays | | 3 Delays | | 4 Delays | |
|---|---|---|---|---|---|---|---|---|---|
| | Inputs | MSE | Correlation | MSE | Correlation | MSE | Correlation | MSE | Correlation |
| 6 Nodes | $L_{s,t}$ | 0.0053 | 0.9885 | 0.0049 | 0.9893 | 0.0046 | 0.9899 | 0.0050 | 0.9892 |
| | $L_{s,t}, \tau_{t-1}, L_{s,t-1}$ | 0.0050 | 0.9890 | 0.0052 | 0.9891 | 0.0050 | 0.9892 | 0.0054 | 0.9884 |
| | $L_{s,t}, \tau_{t-1}, L_{s,t-1}, \tau_{t-2}, L_{s,t-2}$ | 0.0056 | 0.9878 | 0.0052 | 0.9886 | 0.0050 | 0.9891 | 0.0050 | 0.9890 |
| 8 Nodes | $L_{s,t}$ | 0.0053 | 0.9885 | 0.0050 | 0.9891 | 0.0050 | 0.9892 | 0.0050 | 0.9893 |
| | $L_{s,t}, \tau_{t-1}, L_{s,t-1}$ | 0.0050 | 0.9891 | 0.0050 | 0.9891 | 0.0047 | 0.9898 | 0.0048 | 0.9896 |
| | $L_{s,t}, \tau_{t-1}, L_{s,t-1}, \tau_{t-2}, L_{s,t-2}$ | 0.0048 | 0.9896 | 0.0047 | 0.9901 | 0.0050 | 0.9891 | 0.0048 | 0.9896 |
| 10 Nodes | $L_{s,t}$ | 0.0058 | 0.9877 | 0.0051 | 0.9891 | 0.0049 | 0.9894 | 0.0051 | 0.9889 |
| | $L_{s,t}, \tau_{t-1}, L_{s,t-1}$ | 0.0047 | 0.9897 | 0.0048 | 0.9895 | 0.0047 | 0.9897 | 0.0048 | 0.9895 |
| | $L_{s,t}, \tau_{t-1}, L_{s,t-1}, \tau_{t-2}, L_{s,t-2}$ | 0.0043 | 0.9907 | 0.0052 | 0.9889 | 0.0048 | 0.9895 | 0.0045 | 0.9903 |
| 12 Nodes | $L_{s,t}$ | 0.0053 | 0.9884 | 0.0059 | 0.9877 | 0.0046 | 0.9900 | 0.0053 | 0.9884 |
| | $L_{s,t}, \tau_{t-1}, L_{s,t-1}$ | 0.0046 | 0.9900 | 0.0047 | 0.9898 | 0.0044 | 0.9904 | 0.0050 | 0.9895 |
| | $L_{s,t}, \tau_{t-1}, L_{s,t-1}, \tau_{t-2}, L_{s,t-2}$ | 0.0052 | 0.9888 | 0.0041 | 0.9911 | 0.0044 | 0.9907 | 0.0044 | 0.9905 |
| 14 Nodes | $L_{s,t}$ | 0.0057 | 0.9876 | 0.0044 | 0.9905 | 0.0045 | 0.9902 | 0.0062 | 0.9867 |
| | $L_{s,t}, \tau_{t-1}, L_{s,t-1}$ | 0.0046 | 0.9900 | 0.0051 | 0.9889 | 0.0042 | 0.9910 | 0.0050 | 0.9894 |
| | $L_{s,t}, \tau_{t-1}, L_{s,t-1}, \tau_{t-2}, L_{s,t-2}$ | 0.0069 | 0.9854 | 0.0048 | 0.9895 | 0.0045 | 0.9903 | 0.0046 | 0.9902 |
| 16 Nodes | $L_{s,t}$ | 0.0056 | 0.9876 | 0.0052 | 0.9887 | 0.0048 | 0.9895 | 0.0047 | 0.9897 |
| | $L_{s,t}, \tau_{t-1}, L_{s,t-1}$ | 0.0056 | 0.9883 | 0.0044 | 0.9905 | 0.0052 | 0.9890 | 0.0070 | 0.9847 |
| | $L_{s,t}, \tau_{t-1}, L_{s,t-1}, \tau_{t-2}, L_{s,t-2}$ | 0.0054 | 0.9888 | 0.0050 | 0.9895 | 0.0044 | 0.9906 | 0.0053 | 0.9885 |
| 18 Nodes | $L_{s,t}$ | 0.0051 | 0.9889 | 0.0049 | 0.9894 | 0.0042 | 0.9910 | 0.0051 | 0.9903 |
| | $L_{s,t}, \tau_{t-1}, L_{s,t-1}$ | 0.0047 | 0.9897 | 0.0047 | 0.9898 | 0.0047 | 0.9899 | 0.0046 | 0.9901 |
| | $L_{s,t}, \tau_{t-1}, L_{s,t-1}, \tau_{t-2}, L_{s,t-2}$ | 0.0046 | 0.9899 | 0.0054 | 0.9883 | 0.0048 | 0.9899 | 0.0044 | 0.9905 |
| 20 Nodes | $L_{s,t}$ | 0.0055 | 0.9880 | 0.0045 | 0.9903 | 0.0054 | 0.9883 | 0.0053 | 0.9884 |
| | $L_{s,t}, \tau_{t-1}, L_{s,t-1}$ | 0.0045 | 0.9901 | 0.0047 | 0.9897 | 0.0073 | 0.9850 | 0.0163 | 0.9771 |
| | $L_{s,t}, \tau_{t-1}, L_{s,t-1}, \tau_{t-2}, L_{s,t-2}$ | 0.0039 | 0.9916 | 0.0046 | 0.9901 | 0.0055 | 0.9882 | 0.0052 | 0.9888 |

# Appendix C: Standard Network Results for Opportunity

| | Standard Network Results for Opportunity | | | | | |
|---|---|---|---|---|---|---|
| Inputs | $L_{s,t}$ | | $L_{s,t}, \tau_{t-1}, L_{s,t-1}$ | | $L_{s,t}, \tau_{t-1}, L_{s,t-1}, \tau_{t-2}, L_{s,t-2}$ | |
| Number of Nodes | MSE | Correlation | MSE | Correlation | MSE | Correlation |
| 6 Nodes | 0.0764 | 0.6231 | 0.0035 | 0.9860 | 0.0034 | 0.9864 |
| 8 Nodes | 0.0763 | 0.6240 | 0.0035 | 0.9859 | 0.0045 | 0.9819 |
| 10 Nodes | 0.0755 | 0.6291 | 0.0038 | 0.9850 | 0.0033 | 0.9869 |
| 12 Nodes | 0.0757 | 0.6280 | 0.0036 | 0.9856 | 0.0031 | 0.9875 |
| 14 Nodes | 0.0788 | 0.6081 | 0.0034 | 0.9864 | 0.0034 | 0.9864 |
| 16 Nodes | 0.0751 | 0.6316 | 0.0035 | 0.9860 | 0.0031 | 0.9877 |
| 18 Nodes | 0.0756 | 0.6287 | 0.0036 | 0.9855 | 0.0032 | 0.9874 |
| 20 Nodes | 0.0751 | 0.6312 | 0.0032 | 0.9871 | 0.0032 | 0.9874 |

# Appendix D: Standard Network Results for Spirit

| | Standard Network Results for Spirit | | | | | |
|---|---|---|---|---|---|---|
| Inputs | $L_{s,t}$ | | $L_{s,t}, \tau_{t-1}, L_{s,t-1}$ | | $L_{s,t}, \tau_{t-1}, L_{s,t-1}, \tau_{t-2}, L_{s,t-2}$ | |
| Number of Nodes | MSE | Correlation | MSE | Correlation | MSE | Correlation |
| 6 Nodes | 0.1368 | 0.6376 | 0.0053 | 0.9885 | 0.0052 | 0.9887 |
| 8 Nodes | 0.1297 | 0.6612 | 0.0053 | 0.9884 | 0.0049 | 0.9894 |
| 10 Nodes | 0.1227 | 0.6835 | 0.0051 | 0.9890 | 0.0043 | 0.9907 |
| 12 Nodes | 0.1255 | 0.6761 | 0.0053 | 0.9884 | 0.0045 | 0.9902 |
| 14 Nodes | 0.1287 | 0.6636 | 0.0054 | 0.9883 | 0.0067 | 0.9853 |
| 16 Nodes | 0.1220 | 0.6854 | 0.0050 | 0.9891 | 0.0050 | 0.9896 |
| 18 Nodes | 0.1236 | 0.6808 | 0.0052 | 0.9888 | 0.0043 | 0.9906 |
| 20 Nodes | 0.1242 | 0.6805 | 0.0052 | 0.9888 | 0.0048 | 0.9897 |